\documentclass[aps,prl,twocolumn,byrevtex]{revtex4-1}
\usepackage{times}% for making `pdf' file
\usepackage{mathptmx}
\usepackage{graphicx}
\usepackage{hyperref}
\usepackage{amsfonts}
\usepackage{amsmath}
\usepackage{amssymb}
\usepackage{natmove}

\begin{document}
\title{Reentrant Quantum Spin Hall States in Charge Density Wave Phase of Doped Single-Layer Transition Metal Dichalcogenides}

\author{Jun-Ho Lee}
\author{Young-Woo Son}
\email[Email: ]{hand@kias.re.kr}
\affiliation{Korea Institute for Advanced Study, 85 Hoegiro, Seoul 02455, Korea}

\date{\today}

\begin{abstract}
Using first-principles calculation methods, we reveal a series of phase transitions as a function of electron doping in single-layer 1T$'$-MoTe$_2$ and 1T$'$-WTe$_2$ exhibiting quantum spin Hall (QSH) edge states without doping. As increasing doping, we show that a phonon mediated superconducting phase first realizes and is followed by a charge density wave (CDW) phase with a nonsymmorphic lattice symmetry. The newly found CDW phase exhibits Dirac or Weyl energy bands with a spin-orbit coupling in case of a fractional band filling and re-enters into topological insulating phase with fully filled bands. The robust resurgence of QSH state coexisting with the CDW phase is shown to originate from band inversions induced by the nonsymmorphic lattice distortion through the strong electron-phonon interaction, thus suggesting a realization of various interfacial states between superconducting, density wave and topological states on a two-dimensional crystal only by doping.
\end{abstract}

\maketitle

%\section{Introduction}
Layered transition metal dichalcogenides (TMDs) have shown plentiful collective phenomena as well as topological electronic properties. Prominent examples are charge density wave (CDW)~\cite{wil}, superconducting (SC) phase~\cite{wil2}, two-dimensional (2D) quantum spin Hall (QSH) state~\cite{qia}, and three-dimensional (3D) Weyl semimetal (WSM)~\cite{sol} to name a few. Among these, CDW and SC in bulk TMDs have long been studied over several decades~\cite{wil,wil2}. Recently, there are intensive efforts to understand new physical properties shown in a single-layer limit of TMDs that may differ from those in their bulk form~\cite{cal,xi,uge,iwasa_review}.

Maintaining their stoichiometry, TMDs have several polymorphic structures showing radically different electronic properties~\cite{chh}. The most common atomic structures are the trigonal prismatic (2H) and octahedral (1T) form. For single-layer MoTe$_2$ and WTe$_2$, the 1T structure becomes to be unstable and turns into the distorted octahedral structure called as 1T$'$ form as shown in Fig. 1(a)~\cite{keum,qia}. 
One of the most fascinating phenomena shown in 1T$'$ structures is the emergence of topologically nontrivial states. For example, 2D QSH state in a single-layer 1T$'$ structure is theoretically predicted~\cite{qia} and experimentally confirmed recently~\cite{fei,tan,tan2}. The stacked 1T$'$ layer with the orthorhombic structure is known as a type-II WSM, while one with the monoclinic structure is a trivial metal~\cite{sol,kim2,sun}.

Aforementioned distinct states could be enhanced, intertwined or coexist by external perturbations such as chemical or carrier doping~\cite{mor,rosner,yu,jtye,lli,yge}. Doping-induced phase transitions between different polymorphic structures for 2H and 1T forms have been studied theoretically~\cite{duerloo,kan} and experimentally~\cite{wan,lin2} while those for single-layer 1T$'$ structures have not been studied well.
Since the 1T$'$-TMDs have shown various topological states differing from 2H and 1T forms, there would be an interesting interplay between carrier dopings, structural phase transition, collective phenomena as well as topological states. Motivated by rapid developments in this field realizing extremely thin samples~\cite{fei,tan,tan2,wan}, we have performed a comprehensive theoretical study on possible phase transition as a function of doping in single-layer 1T$'$-TMDs.

\begin{figure}[b]
    \includegraphics[width=1.0\columnwidth]{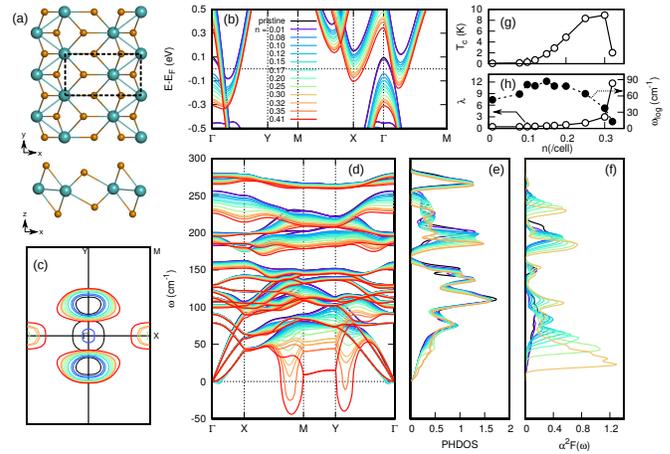}
	\caption{(a) Top (upper) and side (lower) view of optimized structure of single-layer 1T$'$-MoTe$_2$. Large green (small orange) circles represent Mo (Te) atoms. Dotted lines represent unitcell. (b) Electronic band structures along high-symmetry lines, (c) Fermi surfaces, (d) phonon dispersions, (e) phonon density of states, and (f) Eliashberg spectral functions of 1T$'$-MoTe$_2$ with different doping ($n$). In (b), colored lines are denoted for different $n$ (see its definition in the text). (g) Transition temperature of superconducting phase ($T_{\rm c}$) as a function of $n$. (h) Open and filled circles represent electron-phonon coupling strength ($\lambda$) and logarithmic averages of phonon frequencies ($\omega_{\rm log}$) as a function of $n$.}
\end{figure}

In this Letter, using various first-principles calculation methods, we explore a series of phase transitions in single-layer 1T$'$-MoTe$_2$ and 1T$'$-WTe$_2$ (henceforth MoTe$_2$ and WTe$_2$ for simplicity) as a function of electron doping. We find that the phase transitions in the 1T$'$-TMDs quite differ from those shown in other TMDs. Without doping, these single-layer TMDs shows QSH edge states~\cite{qia}. With a low doping, a phonon-mediated SC phase with a maximum transition temperature ($T\rm_c$) of 9 K is shown to be realizable. As increasing doping, we find that SC phase fades away and that a new CDW phase emerges. The crystal structures in the newly found CDW phase are distorted in such a way to respect the nonsymmorphic lattice symmetry so that 2D Dirac or Weyl semimetallic energy bands with spin-orbit coupling~\cite{you} are made possible. We also find that a topological QSH state appears again when the energy bands of the new CDW phase are fully filled. So, the coexistence of QSH state and CDW phase are predicted.  
With extremely higher doping, the lattice is distorted further with a significant band gap and without topological states. Since the doping concentrations considered in this study are already achieved in recent experiments~\cite{jtye,lli,yu}, we believe that various interesting interfacial structures between SC phase, CDW states and topological insulator~\cite{fu1,fu2,har,pri} can be manipulated on the single layer 1T$'$-TMDs by external doping. 

%\section{Method}
Optimized lattice structures of doped systems are obtained through first-principles calculation methods using Quantum ESPRESSO~\cite{gia} and Vienna Ab-initio Simulation Package (VASP)~\cite{vasp}. We use the PBE generalized gradient approximation~\cite{pbe} for the exchange-correlation functional and consider the on-site Coulomb repulsion ($U$) for the specific cases. After obtaining the ground lattice structures with doping, we compute electron-phonon couplings and solve the Eliashberg equation using the EPW code~\cite{pon}. To calculate $\mathbb{Z}_2$ invariant with and without inversion symmetry, we used Z2Pack~\cite{z2pack1,z2pack2}. The electron doping is simulated by adding the electron with the opposite background charge as well as by alkali metal adsorptions. The detailed computational methods can be found in supplemental materials~\cite{supple}. 

%\section{Results}
First, we compute optimized lattice structures with increasing doping concentration ($n$) up to 0.41 electrons per a unit cell of 1T$'$ structure (denoted as $n=0.41/\rm cell$) corresponding to  2D carrier density ($n_{\rm 2D}$) of 1.80 $\times$ 10$^{14}$/cm$^2$. The optimized lattice constants increase with increasing $n$ [Fig. S1 in \cite{supple}]. Figure 1(b) shows electronic band structures of MoTe$_2$ with varying $n$. Fermi energy ($E\rm _F$) rigidly shifts up by electron doping while the corresponding Fermi surface (FS) changes a lot. Our computed FS is consistent with a recent ARPES experiment~\cite{tan2}. Figure 1(c) displays FSs with increasing $n$. The size of hole pocket at $\Gamma$ point decreases with doping and eventually disappears when $n > 0.05/\rm cell$. The area of electron pocket located on $\Gamma-$Y line increases with doping and an electron pocket emerges at X point when $n> 0.24/\rm cell$. {\bf }

Figures 1(d)-(f) represent phonon dispersions, phonon density of states (PHDOS), and Eliashberg spectral function $\alpha^2F(\omega)$ of MoTe$_2$ with varying $n$, respectively. Because of the increase of lattice constants with doping, overall phonon frequencies tend to decrease as increasing $n$. At ${\bf q }$ $\simeq$ $4/5 \rm XM$ and $4/5\rm \Gamma Y$, the low energy phonons soften significantly with doping shown in Fig. 1(d). So, the PHDOS around 50 $\rm cm^{-1}$ slightly increases and $\alpha^2F(\omega)$ strongly enhances, implying that the low energy phonon modes contribute SC phase of doped MoTe$_2$. This agrees with recent studies on SC in other single-layer TMDs~\cite{rosner,hei}. The calculated $T\rm_c$ of SC phase using Allen-Dynes formula, electron-phonon coupling constant ($\lambda$), and logarithmic average of phonon frequencies ($\omega_{\rm log}$) as a function of $n$ are plotted in Fig. 1(g) and 1(h), respectively. As $n$ increases, $\lambda$ increases and $\omega_{\rm log}$ decreases, thus showing the dome-shaped $T$-$n$ diagram. The maximum $T\rm_c$ of MoTe$_2$ reaches $\sim$ 9 K at $n= 0.30/\rm cell$ ($n_{2D}\simeq1.32 \times 10^{14} \rm /cm^2$). WTe$_2$ shares almost similar properties with MoTe$_2$ and its maximum $T\rm_c$ is $\sim$ 7 K at $n= 0.20/\rm cell$ as shown in Fig. S2. 

\begin{figure}[t]
	\includegraphics[width=1.0\columnwidth]{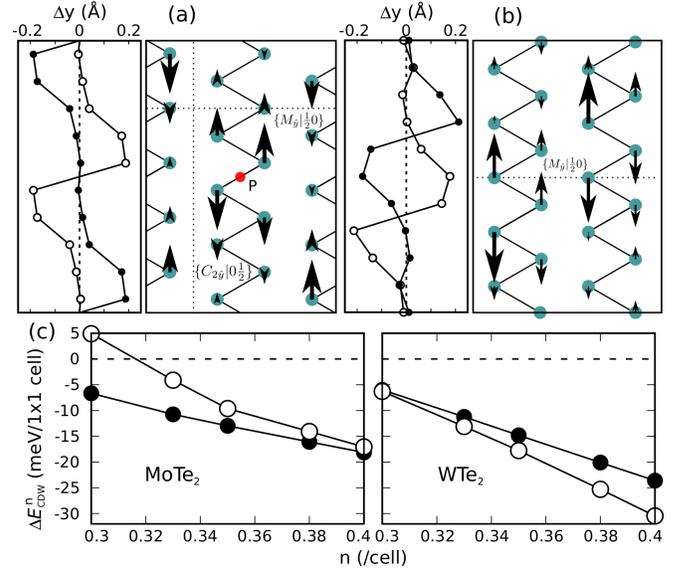}
	\caption{Atomic displacement along y-axis ($\Delta$y) (left panel) and its direction (right panel) of transition metals (TMs) in the 2 $\times$ 5 CDW (a) with and (b) without inversion symmetry, respectively. The green circles forming zigzag chains represent TMs. The orange circles for Te atoms in Fig. 1(a) do not show here for clarity. The open and filled circles represent two different zigzag chains of TMs. The red circle with P in (a) denotes the inversion center. The dotted lines show glide mirror plane $M_{\hat{y}}$ and screw axis $C_{2\hat{y}}$. (c) The formation energy ($\Delta E^n_{\rm CDW}$) of CDW phases for MoTe$_2$ and WTe$_2$, respectively. Filled (open) circles represent $\Delta E^n_{\rm CDW}$ for 2 $\times$ 5 CDW with (without) inversion.}
\end{figure}

With further doping, the atomic structure of the pristine 1T$'$-TMDs becomes to be unstable. When $n\geq 0.32 (0.22) /\rm cell$ corresponding to $n_{2D}\geq 1.40 (0.97) \times 10^{14}/\rm cm^2$ for MoTe$_2$ (WTe$_2$), the frequencies of the softened acoustic phonon mode are negative around ${\bf q}_{\rm CDW}\simeq 4/5$XM as shown in Fig. 1(d). The lattice instability is relieved by forming a new stable structure with a $2\times5$ supercell with respect to the unit cell (called as $2\times5$ CDW hereafter) where the transition metals along with zigzag chain direction are distorted as shown in Fig. 2(a). The pattern of lattice distortions retains inversion symmetry and shows antisymmetric sinusoidal wave with respect to inversion center as shown in Fig. 2(a). We find that the largest atomic displacement of transition metals is $5.3\%$ of the lattice constant and that the ${\bf q}_{\rm CDW}$ does not depend on doping or on changes of FS with doping. Thus, like other TMDs~\cite{cal,ros,joh}, the $2\times5$ CDW phase is generated by the strong electron-phonon coupling rather than by the simple nesting of FSs. 
We also find another CDW phase in the $2\times5$ supercell where the pattern of lattice distortion breaks inversion symmetry as shown in Fig. 2(b). The detailed structural information can be found in Fig. S3.
It is noticeable that the both new phases respect one or two kinds of nonsymmorphic crystal symmetries regardless of  the presence of inversion symmetry as will be discussed later.

\begin{figure}[t]
	\includegraphics[width=1.0\columnwidth]{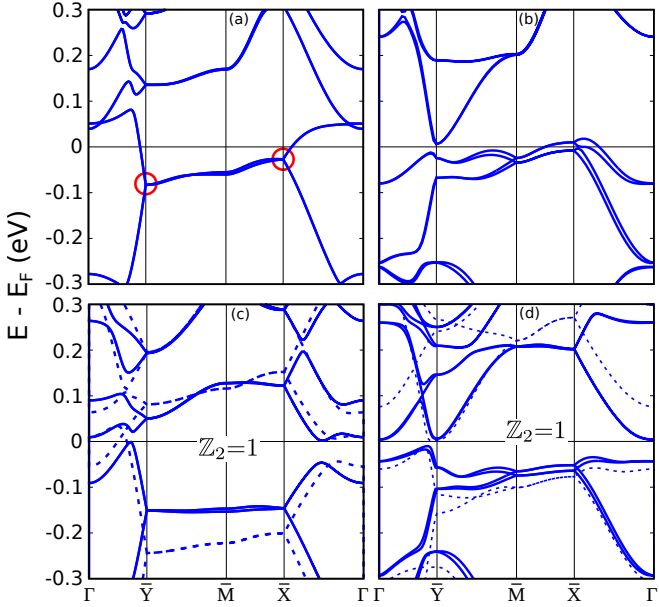}
	\caption{Electronic band structures of the $2\times5$ CDW of doped WTe$_2$  (a) with inversion ($n=0.28/$cell) and (b) without inversion symmetry ($n=0.35/$cell). The BZ corresponding to the $2\times5$ CDW can be found in Fig. S8. The fourfold-degenerated Dirac points are depicted by red circles in (a). Electronic band structures of the $2\times5$ CDW of doped MoTe$_2$ ($n=0.40/$cell) (c) with and (d) without inversion symmetry. The dashed lines in (c) and (d) show band structures without spin-orbit coupling. }
\end{figure}

% energetics for doping concentrations
We investigate the stability of $2\times5$ CDW phases by calculating its formation energy ($\Delta E^n_{\rm CDW}$) 
as a function of doping $n$ for $0.3/{\rm cell} \leq n \leq 0.4 /\rm cell$ where
 $\Delta E^n_{\rm CDW} \equiv (E^n_{i\times j}-M\cdot E^n_{1\times1})/M$. 
 Here, $E^n_{i\times j}$ ($E^n_{1\times1}$) stands for the total energy of $i\times j$ CDW phase ($1\times1$ 1T$'$ unitcell) for a given doping of $n$ and $M\equiv i\times j$, respectively. 
Figure 2(c) displays $\Delta E^n_{\rm CDW}$ for MoTe$_2$ and WTe$_2$. 
For MoTe$_2$ (WTe$_2$), the $2\times5$ CDW phase with (without) the inversion symmetry [Fig. 2(a)] is more stable than one without (with) the inversion [Fig. 2(b)] for all $n$ in the range while the energy difference between them decreases (increases) with increasing $n$. 
It is also noteworthy that $\Delta E^n_{\rm CDW}$ of WTe$_2$ is larger than that of MoTe$_2$,
implying the enhanced stability as well as a higher CDW transition temperature of WTe$_2$ than those of MoTe$_2$.
We extensively search other possible supercell structures and find few stable structures such as a CDW phase with a $2\times8$ supercell [Fig. S4]. However, all other structures have higher $\Delta E^n_{\rm CDW}$ than one of the $2\times5$ CDW, thus confirming that the new phases are quite stable for a wide range of $n$.

The new CDW phase of doped 1T$'$-TMDs show interesting topological electronic properties owing to their crystal symmetries.  The crystal structure of $2\times5$ CDW with the inversion symmetry has two nonsymmorphic crystal symmetries, \{$M_{\hat{y}}|\frac{1}{2}0$\} and \{$C_{2\hat{y}}|0\frac{1}{2}$\} as shown in Fig. 2(a). Thanks to the symmetry, with a fractional band filling, the metallic energy bands in the phase must have the fourfold-degenerated Dirac points at the $\overline{\rm X}$ and $\overline{\rm Y}$ points~\cite{you} as shown in Figs. 3(a) (Red circles). The similar structure without the inversion [Fig. 2(b)] also has the nonsymmorphic symmetry of \{$M_{\hat{y}}|\frac{1}{2}0$\}, thus showing the twofold-degenerated Weyl points along $M_{\hat{y}}$ invariant lines at $k_y =$ 0 and $\pi$ as shown in Figs. 3(b). 
Dirac cone located at $\overline{\rm Y}$ ($\overline{\rm X}$) point exhibits strong anisotropy such that the Fermi velocity, $v_{\rm F}=\frac{1}{\hbar}\frac{\partial E}{\partial \bf k}$, at $\overline{\rm Y}$ ($\overline{\rm X}$) pointing along $\overline{\rm Y}$$\Gamma$ ($\overline{\rm X}$$\Gamma$) direction is 11.5 (3.67) $\times$ 10$^5$ m$/$s, while $v_{\rm F}$ along $\overline{\rm Y}$$\overline{\rm M}$ ($\overline{\rm X}$$\overline{\rm M}$) direction is 6.47 (19.4) $\times$ 10$^3$ m$/$s, respectively. From the calculated band structures, we can infer that 1T$'$-TMD with a strong spin-orbit coupling (SOC) is a quite unique material showing both 2D Dirac or Weyl in its doped single layer and 3D Weyl fermions in its stacked structures, respectively. 

The new $2\times5$ CDW phase can coexist with the QSH insulating phase when the CDW energy bands are fully filled. Figure 3(c) shows the electronic band structure of the CDW phase with the inversion symmetry in case of $n=0.40/\rm cell$.  It is shown that without considering SOC the CDW energy gap ($E_g$) opens owing to the lattice distortion except the $\Gamma$$\overline{\rm Y}$ line where the Dirac cone by the band inversion develops. Including SOC, the spin-orbit gap of 22 meV opens at the Dirac point as shown in Fig. 3(c). The band structure of $2\times5$ CDW without the inversion symmetry and with the SOC show the lifted Kramers degeneracy with an indirect $E_g$ of 14 meV, as shown in Fig. 3(d). In order to investigate their topological properties, we compute the evolution of hybrid Wannier charge centers as shown in Fig. S5 and obtain the odd $\mathbb{Z}_2$ invariant, indicating that the both structures are topologically nontrivial. 
Since the calculated energy band gaps and band widths of single-layer TMDs match experiment results only with proper considerations of Coulomb interactions~\cite{keum,uge,kim2,sjkim}, we add the $U$ and find the $E_g$ increases with increasing $U$. For example, the addition of $U$ of 4 eV on the transition metals increases $E_g$ to 65 (80) meV for 2 $\times$ 5 CDW with inversion of MoTe$_2$ (WTe$_2$).

\begin{figure}[t]
	\includegraphics[width=1.0\columnwidth]{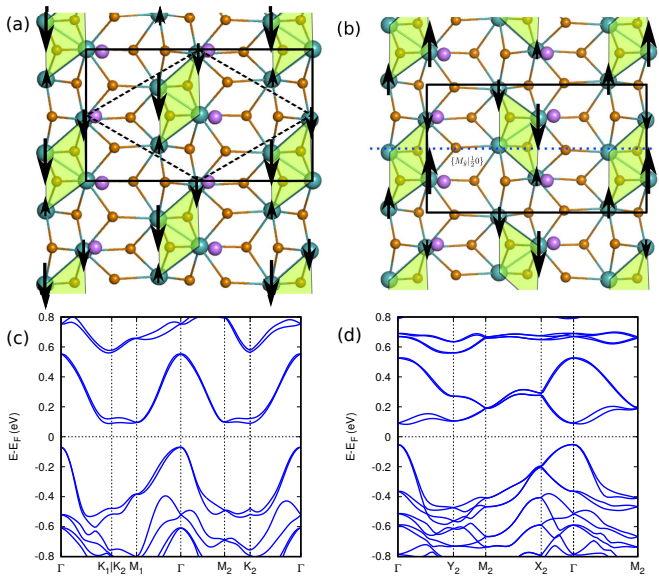}
	\caption{The optimized structures of (a) D11 and (b) D22 of the $2\times2$ CDW in Li doped MoTe$_2$. Solid lines represent $2\times2$ supercell. In (a), rhombus shape drawn by dashed lines represents primitive cell of D11. The direction of arrows on TMs represents the atomic displacements and their sizes are proportional to displacement magnitude. In (b), glide mirror plane $M_{\hat{y}}$ is marked by blue dotted line.  Shaded regions represent the diamond chain structure. Electronic band structure of (c) D11 and (d) D22 along high-symmetry points of their primitive cells, respectively. The BZ corresponding to the $2\times2$ CDW can be found in Fig. S8.}
\end{figure}

For a higher doping case of $n>0.4/\rm cell$, we consider adsorptions of hydrogen and alkali metal atoms on the systems following recent experiments demonstrating extreme high doping in TMDs ($n_{2D}> 10^{15}/\rm cm^2$) through Li intercalation~\cite{yu} or hydrogen adsorption~\cite{2dm}. 
We use H, Li, Na, and K atoms for doping. Here we only consider a full coverage of adatoms (one adatom per $1\times1$ unitcell of 1T$'$ structure) to simulate an extremely higher doping. A comparison between electronic band structures of Li doped MoTe$_2$ and pristine MoTe$_2$ is shown in Fig. S6(a). It is clearly demonstrated that the $E\rm _F$ is much higher in energy than one considered for reentered QSH phase shown in Fig. 1(b). The topology of FS is also drastically modified compare to those obtained in $2\times5$ CDW phase. It consists of quasi-1D-like bands along $k_x$ direction and hole pocket around zone center as shown in Fig. S6(b). 
We calculate phonon dispersions for Li atoms doped MoTe$_2$ and find the negative frequencies near $\bf q'$$\rm_{CDW} \simeq$ $\Gamma$M as shown in Fig. S6(c), indicating different CDW phase compared with the $2\times5$ CDW phase.

\begin{table}[b]
	\caption{$\it E$$\rm_g$ (meV) and $\Delta$$\it E$$\rm_{CDW}$ (meV/cell) for H, Li, Na, and K doped 1T$'$-MoTe$_2$ (WTe$_2$), respectively.}
	\begin{ruledtabular}
		\begin{tabular}{ccccc}
			&  H & Li & Na & K \\ \hline
			$E_g$& 121 (372) & 158 (362) & 82 (189) & 70 (68) \\
			$\Delta$$\it E$$\rm_{CDW}$  & $-158 (-227)$ & $-129 (-177)$ & $-147 (-176)$& $-166 (-164)$
		\end{tabular}
	\end{ruledtabular}
\end{table}

We find that the $2\times2$ supercell structure with respect to the unitcell of 1T$'$-TMD (called as $2\times2$ CDW) is dynamically stable for the Li doped (also for the other cases as well) MoTe$_2$ [Fig. 4]. This new CDW structure is nothing but a ground atomic structure that can be found in ReS$_2$~\cite{ton}. This atomic reconstruction is called as the diamond chain (DC) structure referring the shape of the new unit cell in the present $2\times2$ CDW in Fig. 4(a) or one in the Re compounds. We note that the similar structures are also found in other studies computing the structures of doped 2H-polymorphs of TMDs~\cite{kan}. The calculated electronic band structure shows the indirect $E_g$ and spin splitting as shown in Fig. 4(c), giving a trivial topology of band structure.
$E_g$ ranges from 68 meV to 372 meV and the total energy gain of $\Delta E\rm_{CDW}$ after the CDW transition ranges from $-$129 meV$/$cell to $-$227 meV$/$cell depending on the TMDs and adsorbates as listed in Table I. $E_g$ and $\Delta$$\it E$$\rm_{CDW}$ of WTe$_2$ are quite larger than those of MoTe$_2$ (except K-doped case).
We also find that the very large displacement of transition metals in the $2\times2$ CDW and its large band gaps implies the strong electron-phonon coupling as an origin of the CDW~\cite{joh,ros}. 

Like several atypical atomic structures shown in ReS$_2$~\cite{ton}, the DC structures in the new $2\times2$ CDW can have various different orientations. So, we can expect similar boundary states between the domains having the different DC orientations as discussed in the Re compounds~\cite{lin}. In the present $2\times2$ CDW phase, four different configurations are possible and the two lowest energy configurations are shown in Figs. 4(a) and (b), called as D11, and D22, respectively (others are shown in Fig. S7). The most stable configuration is the D11 and $\Delta E\rm_{CDW}$ of the D22 shown in Fig. 4(b) is 20 meV$/$cell higher than one of the D11. The latter structure also holds the same nonsymmorphic symmetry of \{$M_{\hat{y}}|\frac{1}{2}0$\} like the $2\times5$ CDW phase shown in Fig. 2(b). So, it gives rise to Weyl points at $k_y =$ 0 and $\pi$~\cite{you} as shown in Fig. 4(d) but the $E_F$ is higher to approach the points. 

In summary, we find the diverse phase transitions between the QSH insulating, SC phase, and normal insulator in doped single-layer 1T$'$-TMDs as a function of electron doping. The new CDW state in doped MoTe$_2$ and WTe$_2$ is shown to coexist with Dirac$/$Weyl metallic state as well as the reentered QSH insulating phase. Considering various intriguing interfacial states between topological states and superconducting phase such as Majorana fermions~\cite{fu1,fu2,har,pri}, we believe that doping controlled phase transition in single-layer 1T$'$-TMDs demonstrated here could provide a new facile platform toward topological state engineering.

\begin{acknowledgements}
We thank S. Kim, S.-H. Kang and H. Yang for fruitful discussions. Y.-W.S. was supported by National Research Foundation (NRF) grant funded by the Korean government (MSIP) (Grant No.2017R1A5A1014862, SRC program: vdWMRC center). The computing resources were supported by the Center for Advanced Computation (CAC) of KIAS.
\end{acknowledgements}

                  %%%%%  REFERENCES  %%%%%

\end{document}

\title{Supplemental material for ``Reentrant Quantum Spin Hall States in Charge Density Wave Phase of Doped Single-Layer Transition Metal Dichalcogenides''}
\author{Jun-Ho Lee}
\author{Young-Woo Son}
\email[Email: ]{hand@kias.re.kr}
\affiliation{Korea Institute for Advanced Study, 85 Hoegiro, Seoul 02455, Korea}

\maketitle

%\tableofcontents

\makeatletter
\renewcommand{\fnum@figure}{\figurename ~{S}\thefigure}
\renewcommand{\fnum@table}{\tablename ~{S}\thetable}
\renewcommand{\theequation}{S\arabic{equation}}
\renewcommand{\bibnumfmt}[1]{[S#1]}
\renewcommand{\citenumfont}[1]{S#1}
\makeatother

\section{Computational details}

We performed density-functional theory calculations by using Quantum ESPRESSO package~\cite{gia}. To calculate Eliashberg equation and electron-phonon coupling (EPC) with fine grids, we used EPW~\cite{pon} package. We used norm-conserving pseudopotential with 80 Ryd energy cutoff and employed the GGA functional of Perdew-Burke-Ernzerhof (PBE)~\cite{pbe} for the exchange-correlation part. The ${\bf k}$-point integration was done by 10 $\times$ 20 $\times$ 1 (5 $\times$ 4 $\times$ 1 and 5 $\times$ 10 $\times$ 1)  Monkhorst-Pack scheme for 1 $\times$ 1 (2 $\times$ 5 and 2 $\times$ 2) cell of 1T$'$ phase. Structural relaxation was performed until the Hellmann-Feynman forces on all the atoms and total energy change became less than 10$^{-4}$ a.u. and 10$^{-6}$ a.u. simultaneously. The vacuum region between adjacent slabs is more than 10 ${\rm \AA}$. In order to simulate the effect of doping, additional electrons are added to the neutral cell with opposite background charge. Dipole correction is considered in the case of a broken inversion symmetry. Methfessel-Paxton scheme for smearing is used with 0.02 Ryd of gaussian spreading. Density-functional perturbation theory (DFPT) is adopted to calculate phonon-dispersion relation with 4 $\times$ 8 $\times$ 1 {\bf q}-point grids with threshold for self-consistency of 10$^{-18}$ Ryd. The electronic wave functions required for the Wannier interpolation within EPW are calculated on uniform and $\Gamma$-centered $\bf k$-point meshes of sizes 4 $\times$ 8 $\times$ 1. For the initial guess of Wannierization, 22 Wannier functions, five $d$-orbitals for two Mo (W) atoms and three $p$-orbitals for four Te atoms, are used to describe the electronic structure. In order to solve the Eliashberg equations, we evaluate electron energies, phonon frequencies, and electron-phonon matrix elements on fine grids using the methods of Ref.~\cite{fel}. The fine grids contain 20 $\times$ 40 $\times$ 1 $\bf k$ and $\bf q$ points. The frequency cutoff ($\omega_c$) is set 1 eV. We calculated the Eliashberg spectral function:

\begin{equation}
\alpha^2F({\omega}) = \frac{1}{2}\int_{\rm BZ}\frac{d\bf q}{\Omega_{\rm BZ}}\omega_{{\bf q}v}\lambda_{{\bf q}v}\delta(\omega-\omega_{{\bf q}v}),
\end{equation}

and the corresponding EPC strength:

\begin{equation}
\lambda = \int_0^{\infty} \frac {\alpha^2 F(\omega)}{\omega}d\omega,
\end{equation} 

where $\lambda_{{\bf q}v}$ and $\omega_{{\bf q}v}$ are the EPC strength and frequency associated with a specific phonon mode $\nu$ and wavevector $\bf q$.  From the calculated $\lambda$, we can estimate transition temperature ($T\rm_c$) of superconducting phase by Allen-Dynes formula: 

\begin{equation}
T\rm_c = \frac{\omega_{log}}{1.2}\exp\Big(-\frac{1.04(1+\lambda)}{\lambda-\mu^*_c(1+0.62\lambda)}\Big),
\end{equation}

where $\rm \omega_{log}$ is a logarithmic average of the phonon frequency, $\rm \mu^*_c$ is the Coulomb pseudopotential with 0.16.
Because unintended free-electron-like (FEL) states appeared in the case of higher doping concentration ($n> 0.41/$cell), we employed the adatom adsorption on the TMDs using VASP~\cite{vasp1,vasp2}. Ion cores are modeled with projector augmented wave (PAW)  potentials~\cite{paw1,paw2}. The semicore $p$ states of K, Na, Mo and W and $s$ state of Li are treated explicitly as valence. For the phonon calculations, supercell and finite displacement approaches, as implemented in phonopy software~\cite{phonopy}, were used with a 2 $\times$ 4 supercell for 1 $\times$ 1 cell (56 atoms) and 2 $\times$ 2 cell (224 atoms) and the atomic displacement distance of 0.01 \AA. 

\newpage

\section{Variation of lattice parameters as a function of doping concentrations}

\begin{figure}[ht]
	\includegraphics[width=17.0cm]{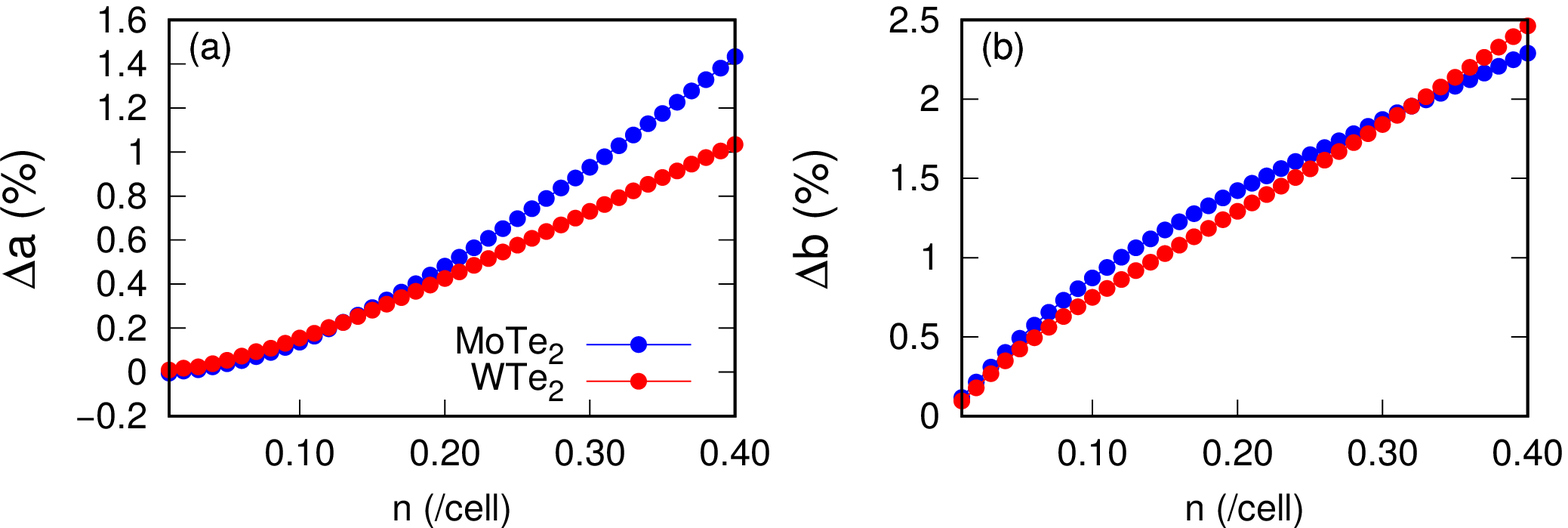}
	\caption{(a) Optimized lattice parameters along $x$- (b) and $y$-axis as a function of $n$ for MoTe$_2$ and WTe$_2$, respectively.}
\end{figure}

\newpage

\section{Electronic, phononic, and superconducting properties of doped 1T$'$-WTe$_2$}

\begin{figure}[ht]
	\includegraphics[width=17cm]{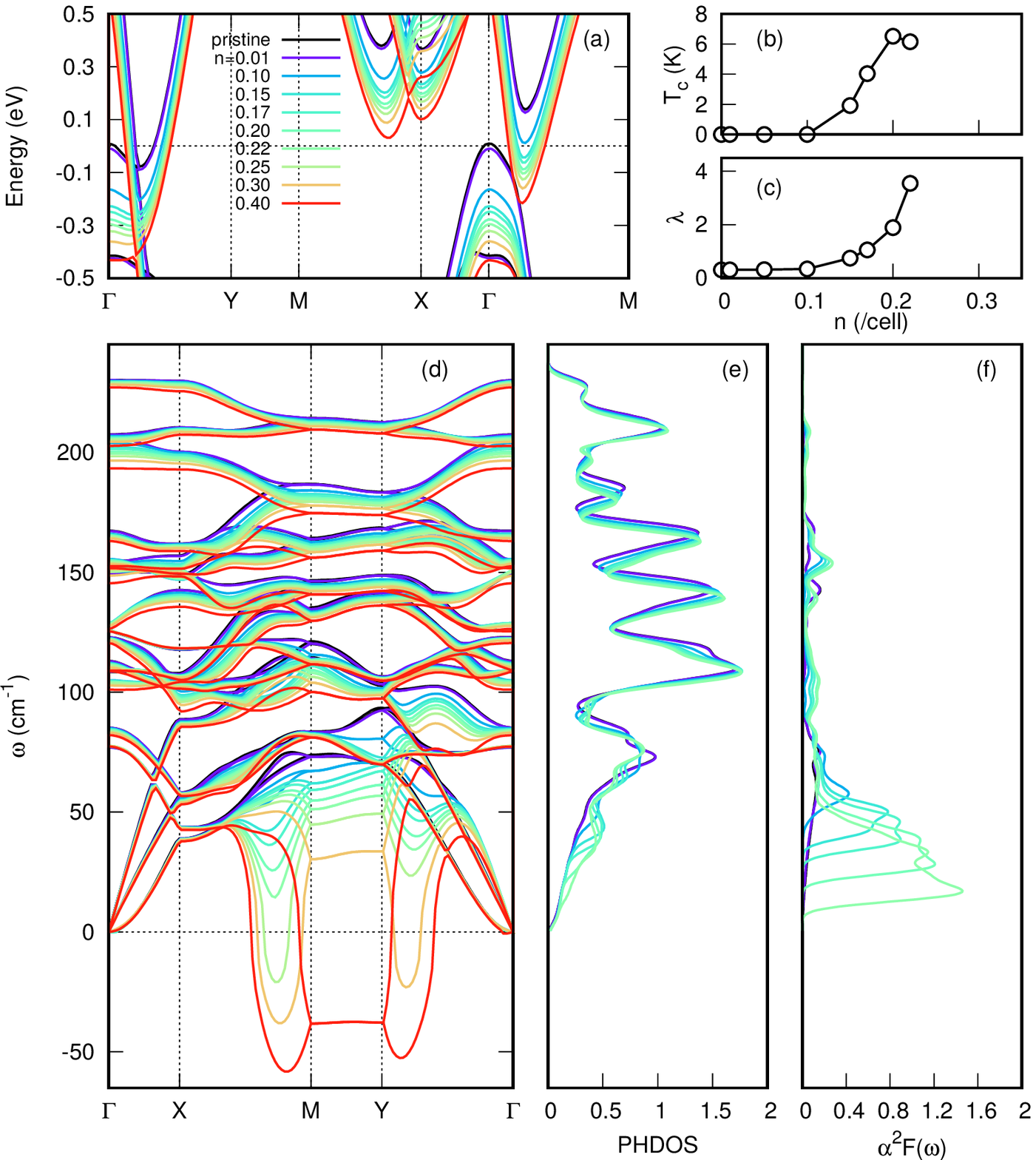}
	\caption{(a) Band structures, (b) transition temperature ($T_{\rm c}$) of superconducting phase, (c) electron phonon coupling strength ($\lambda$), (d) phonon dispersion, (e) phonon density of states (PHDOS), and (f) Eliashberg spectral functions of 1T$'$-WTe$_2$ at different $n$. In (a), zero in energy stands for Fermi level. Each color in (a) denote a different doping of $n/$cell (see text for a definition).}
\end{figure}

\newpage

\section{Atomic structures of 2 $\times$ 5 CDW phase}

\begin{figure}[ht]
	\includegraphics[width=17cm]{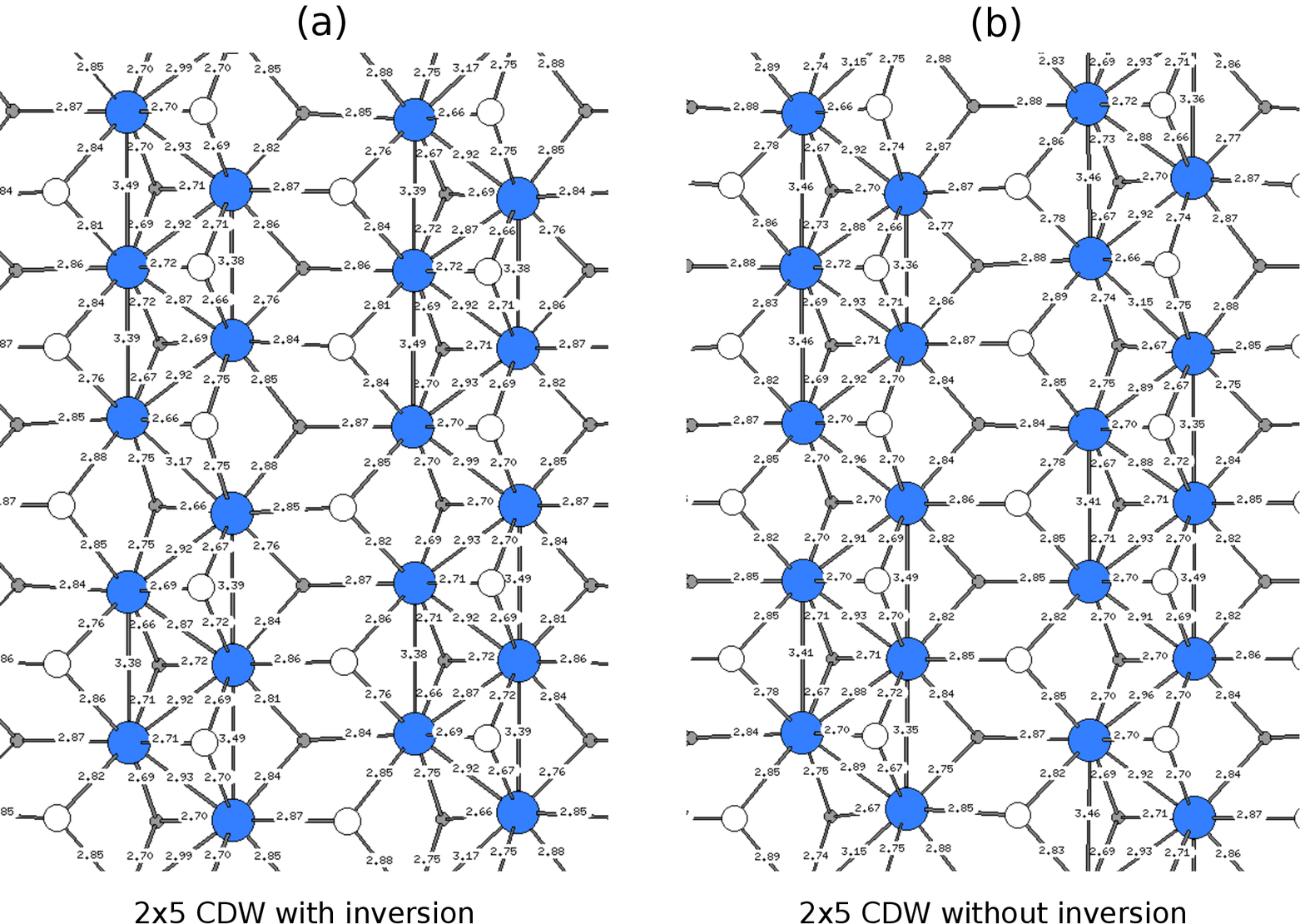}
	\caption{Optimized atomic structures of MoTe$_2$ of 2$\times$ 5 CDW (a) with inversion and (b) without inversion at $n=0.4/$cell with bond length. Blue largest circles, middle size white circles, and smallest gray circles stand for Mo, upper Te, and lower Te atoms, respectively.}
\end{figure}

\newpage

\section{2 $\times$ 8 CDW phase of MoTe$_2$}

We performed structure optimizations in 2 $\times$ 8 supercell which corresponds to $\bf q_{\rm CDW}$ $\simeq$ $ 3/4$XM. The optimized structure has mirror symmetry ($M_{\hat{y}}$) and can be represented by rhombus cell. $\Delta E^n_{\rm CDW}$ for 2 $\times$ 8 CDW are less stable than those obtained in 2 $\times$ 5 CDW. In order to unfold band structure, we used BandUp software~\cite{unfold1,unfold2}.

\begin{figure}[ht]
	\includegraphics[width=17.0cm]{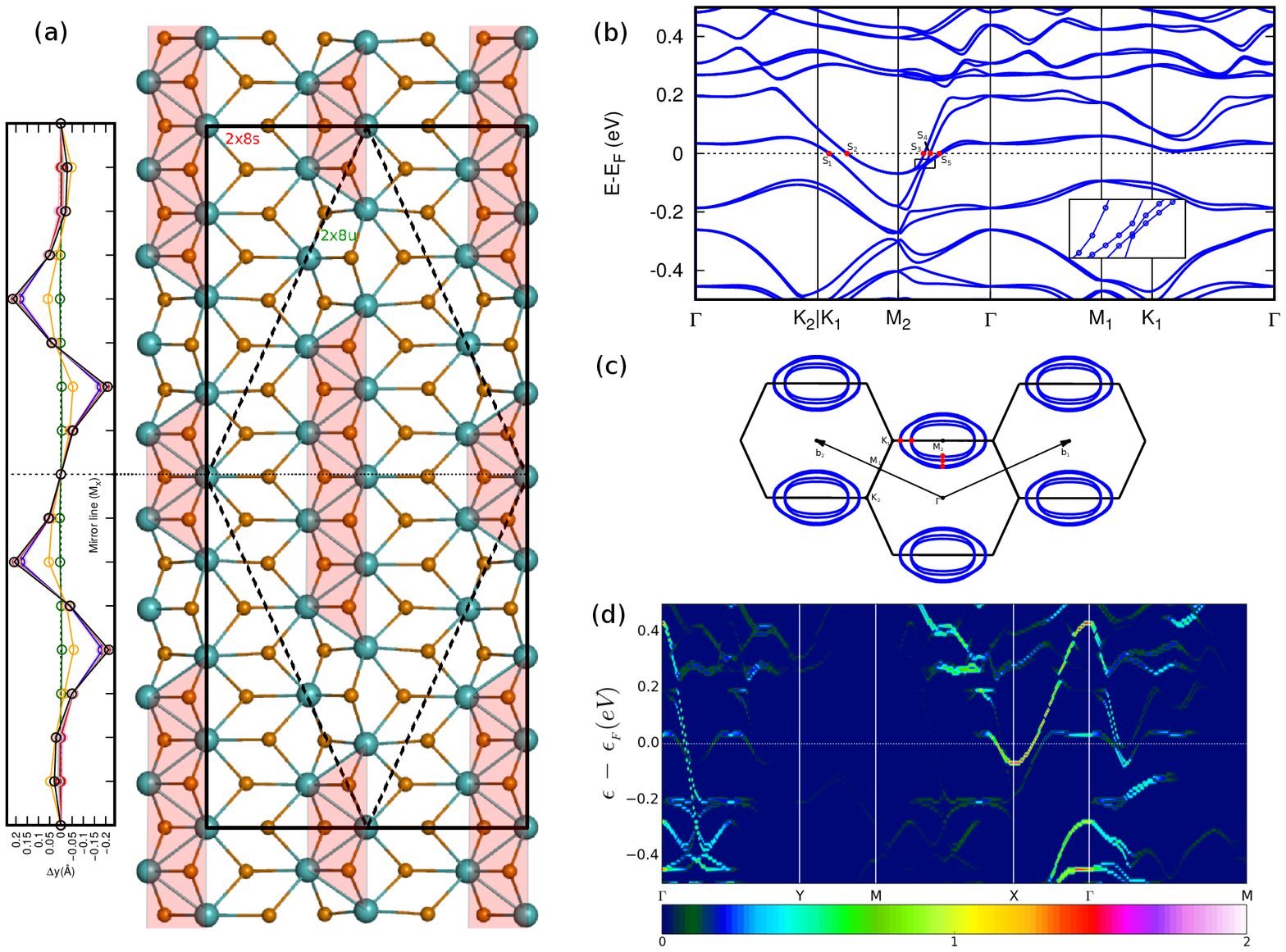}
	\caption{(a) Optimized structure and variation of lattice distortions of 2 $\times$ 8 CDW. The dashed lines forming rhombus represent the primitive cell for 2 $\times$ 8 CDW. (b) Electronic band structure (c) fermi surface and (d) unfolded band structure, respectively.}
\end{figure}

\newpage

\section{Hybrid Wannier charge center}

Figure S5 shows the evolution of hybrid Wannier charge centers (HWCCs) and their largest gap for pristine monolayer 1T$'$-MoTe$_2$ and various CDW phases.

\begin{figure}[ht]
	\includegraphics[width=12.0cm]{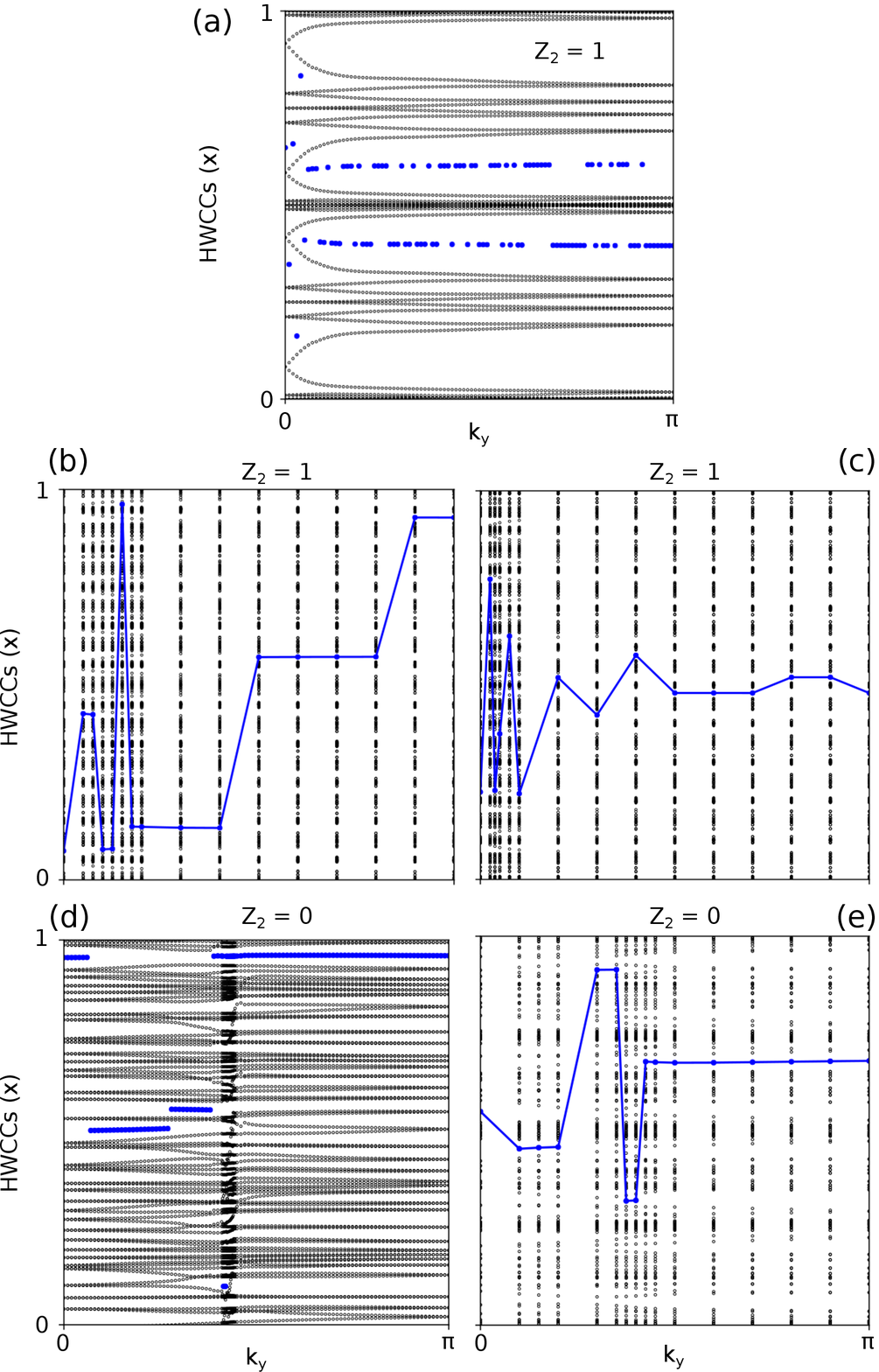}
	\caption{Evolution of hybrid Wannier charge centers (open circles) and their largest gap (blue filled circles) with topological $\mathbb{Z}_2$ invariant at $k_z =0$ for (a) pristine 1T$'$-MoTe$_2$, (b) 2 $\times$ 5 CDW with inversion, (c) 2 $\times$ 5 CDW without inversion, (d) 2 $\times$ 2 CDW with D11 of Li$/$MoTe$_2$, and (e) 2 $\times$ 2 CDW with D21 of Li$/$MoTe$_2$, respectively.}
\end{figure}

\newpage

\section{Adatoms adsorptions on MoTe$_2$ and WTe$_2$}

\begin{figure}[ht]
	\includegraphics[width=16.0cm]{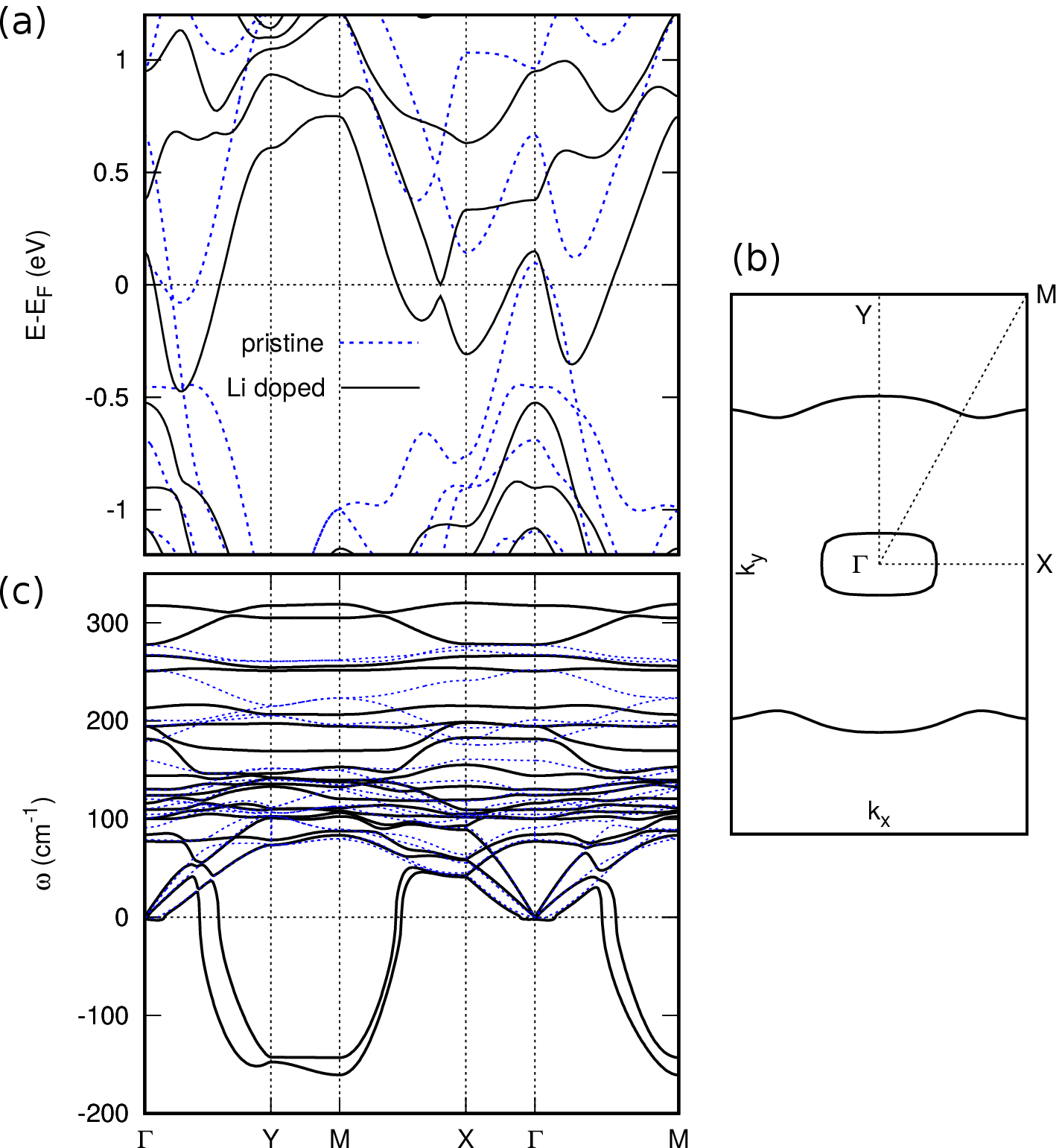}
	\caption{(a) Electronic band structures and (c) phonon dispersions of pristine (blue dotted) and Li doped MoTe$_2$ (black solid), respectively. (b) Fermi surface of Li doped MoTe$_2$.}
\end{figure}

\newpage

\section{Diverse local minima of 2 $\times$ 2 CDW}

\begin{figure}[ht]
	\includegraphics[width=17.0cm]{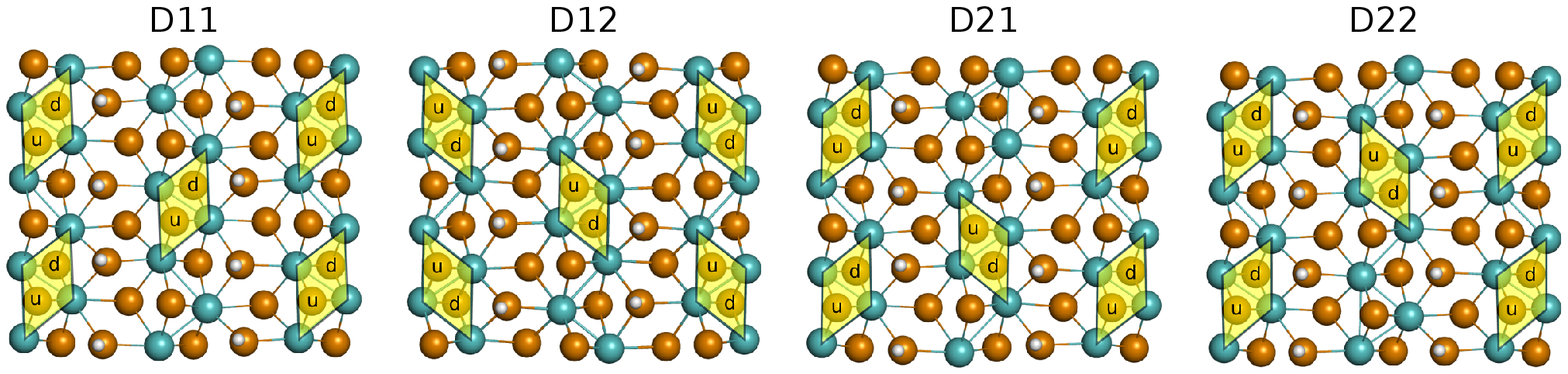}
	\caption{Diverse 2 $\times$ 2 CDW phases. Yellow regions shows diamond shape distortion of transition metals.}
\end{figure}

\newpage

\section{Brillouin zones of various supercells}

\begin{figure}[ht]
	\includegraphics[width=8.0cm]{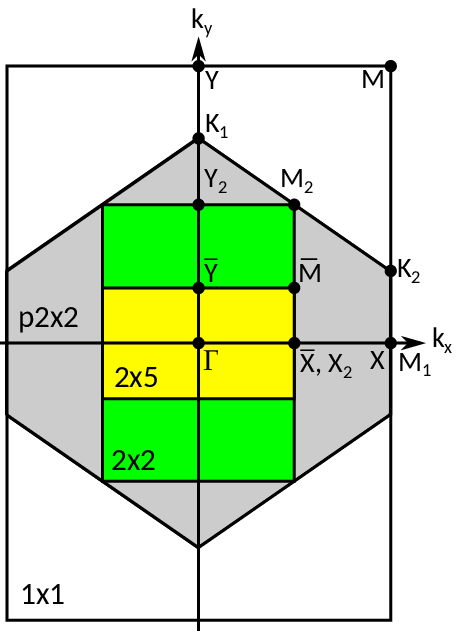}
	\caption{Brillouin zones of 1 $\times$ 1, 2 $\times$ 2 (green), primitive of 2 $\times$ 2 (gray), 2 $\times$ 5 (yellow) supercells and its high-symmetry points, respectively.}
\end{figure}